\documentclass{llncs}
\usepackage{amssymb,amsmath,amscd,latexsym,epic,eepic}
\usepackage{algorithm}
\def\qed{\rule{0.4em}{1.4ex}}

\newcommand{\restr}{\upharpoonright}
\newcommand{\set}[1]{\{#1\}}
\newcommand{\sseq}{\langle s_0,s_1,s_2,\ldots\rangle}

\newcommand{\pat}{\omega}
\newcommand{\Pat}{\Omega}
\newcommand{\straa}{\sigma}
\newcommand{\Straa}{\Sigma}
\newcommand{\strab}{\pi}
\newcommand{\Strab}{\Pi}

\newcommand{\rank}{\mathit{rank}}

\newcommand{\Attr}{\mathit{Attr}}

\newcommand{\Tr}{\mathit{Tr}}
\newcommand{\wh}{\widehat}
\newcommand{\avoidset}{\mathsf{AvoidSetClassical}}

\newcommand{\Inf}{\mathrm{Inf}}

\newcommand{\Reach}[1]{\mathrm{Reach}(#1)}
\newcommand{\Safe}[1]{\mathrm{Safe}(#1)}

\newcommand{\Buchi}[1]{\mathrm{Buchi}(#1)}
\newcommand{\coBuchi}[1]{\mathrm{coBuchi}(#1)}
\newcommand{\RT}{\mathsf{RT}}
\newcommand{\Source}{\mathrm{source}}
\newcommand{\ov}{\overline}

\title{Algorithms for B\"uchi Games\thanks{%
This research was supported in part by
the AFOSR MURI grant F49620-00-1-0327
and the NSF ITR grant CCR-0225610
and the SNSF under the Indo-Swiss Joint Research Programme.}
}
\author{%
Krishnendu Chatterjee\inst{1}%
\and Thomas A.\ Henzinger\inst{1,2}%
\and Nir Piterman\inst{2}
}
\institute{%
University of California, Berkeley, USA\\
\and EPFL, Switzerland \\
{\tt c\_krish@eecs.berkeley.edu, \{tah,Nir.Piterman\}@epfl.ch }
}
\date{}
\begin{document}

\maketitle
\begin{abstract}

The classical algorithm for solving B\"uchi games requires time
$O(n\cdot m)$ for game graphs with $n$ states and $m$ edges.
For game graphs with constant outdegree, the best known algorithm
has running time $O(n^2/\log n)$.
We present two new algorithms for B\"uchi games.
First, we give an algorithm that performs at most $O(m)$ more work
than the classical algorithm, but runs in time $O(n)$ on infinitely
many graphs of constant outdegree on which the classical algorithm
requires time~$O(n^2)$.
Second, we give an algorithm with running time
$O(n\cdot m\cdot\log\delta(n)/\log n)$, where $1\le\delta(n)\le n$
is the outdegree of the game graph.
Note that this algorithm performs asymptotically better than the
classical algorithm if $\delta(n)=O(\log n)$.
\end{abstract}

\section{Introduction}

The algorithmic complexity of solving B\"uchi games is one of the
more intriguing questions in graph algorithms.
The input of the B\"uchi game problem consists of a directed graph 
whose states $S$ are partitioned into
player~1 states and player~2 states, and a set $B\subseteq S$ of
B\"uchi states.
In each player~1 state, the first player chooses an outgoing edge,
and each player~2 state, the second player chooses an outgoing edge.
The question is if, from a given start state, player~1 has a strategy
to visit a state in $B$ infinitely often.

The classical algorithm for solving B\"uchi games has time
complexity $O(n\cdot m)$, where $n$ is the number of states and
$m$ is the number of edges.
The classical algorithm proceeds in a seemingly naive fashion:
first it computes the set $W_1$ of states from which player~1 has
a strategy to visit $B$ once, which requires time~$O(m)$;
then it computes the set of states $\ov{W}_1$ such that player~2
has a strategy to visit $S \setminus W_1$ once.
The set $\ov{W}_1$ is removed from the game graph and the 
algorithm iterates over the reduced game graph unless $\ov{W}_1$ is 
empty for the current game graph.
The algorithm converges in at
most $n$ iterations, is the answer to the B\"uchi game problem.
This algorithm seemingly performs unnecessarily repetitive work,
yet no asymptotically faster algorithm is known.

In~\cite{CJH03}, we gave a subquadratic algorithm for the special case
of graphs with constant outdegree.
In this special case $m=O(n)$, and thus the classical algorithm
performs in time~$O(n^2)$.
We gave an algorithm with running time $O(n^2/\log n)$.
While the classical algorithm computes each set $W_1$ by backward
search of the graph, our algorithm alternated backward searches
with bounded amounts of forward searches.

In this paper, we present two new algorithms for B\"uchi games.
While they fall short of the ultimate goal of a better than
$O(n\cdot m)$ algorithm, they present progress in that direction.
First, in Section~3 we give an algorithm that performs at most
$O(m)$ more work than the classical algorithm.
However, there exist families of game graphs of outdegree 2 on which
our algorithm performs in time~$O(n)$, while the classical algorithm
requires time~$\Omega(n^2)$.
Also there exist families of game graphs of $O(n \log n)$ states with outdegree 
at most 2 on which both the classical and the algorithm of~\cite{CJH03}
requires $O(n^2 \log n)$ time, where as our algorithm requires 
$O(n \log n)$ time.
However, there exist game graphs where our algorithm requires $O(n\cdot m)$ 
time in the worst case.
Our algorithm performs several backward searches, but instead of backward 
search from $B$ it performs backward search from a subset of $S \setminus B$
that are candidates to be in $S\setminus W_1$.

Second, in Section~4 we give an algorithm with running time
$O(n\cdot m\cdot\log\delta(n)/\log n)$, where $1\le\delta(n)\le n$
is the outdegree of the graph.
If $\delta(n)=O(1)$, then this algorithm performs asymptotically
like the algorithm of~\cite{CJH03};
indeed, the new algorithm is based on the algorithm from~\cite{CJH03}.
The forward search of the algorithm of~\cite{CJH03} was very naive;
we develop a more refined forward search which saves repetitive work 
of the forward search of the algorithm of~\cite{CJH03} and thus
obtain the generalization of the algorithm of~\cite{CJH03} to general
game graphs.
If $\delta(n)=O(\log n)$, then our algorithm performs better than
the classical algorithm;
it is to our knowledge the first algorithm that improves on the
classical algorithm in this case.

\section{Definitions}
We consider turn-based deterministic games played by two-players with B\"uchi 
and complementary coB\"uchi objectives for the players, respectively.
We define game graphs, plays, strategies, objectives and notion of winning 
below.

\medskip\noindent{\bf Game graphs.} 
A \emph{game graph} $G=((S,E),(S_1,S_2))$ consists of a directed graph 
$(S,E)$ with a finite state space $S$ and a set $E$ of edges, 
and a partition $(S_1,S_2)$ of the state space $S$ into two sets.
The states in $S_1$ are player~1 states, and the states in $S_2$ are 
player~2 states.
For a state $s\in S$, we write $E(s)=\set{t\in S \mid (s,t) \in E}$ 
for the set of successor states of~$s$.
We assume that every state has at least one out-going edge, 
i.e., $E(s)$ is non-empty for all states $s\in S$.

\smallskip\noindent{\em Plays.}
A game is played by two players: 
player~1 and player~2, who form an
infinite path in the game graph by moving a token along edges.
They start by placing the token on an initial state, and then they
take moves indefinitely in the following way.
If the token is on a state in~$S_1$, then player~1 moves the token along
one of the edges going out of the state.
If the token is on a state in~$S_2$, then player~2 does likewise.
The result is an infinite path in the game graph;
we refer to such infinite paths as plays.
Formally, a \emph{play} is an infinite sequence 
$\sseq$ of states such that $(s_k,s_{k+1}) \in E$ for all $k \geq 0$. 
We write $\Pat$ for the set of all plays.

\smallskip\noindent{\em Strategies.} 
A strategy for a player is a recipe that specifies how to extend plays.
Formally, a \emph{strategy} $\straa$ for player~1 is a function 
$\straa$: $S^* \cdot S_1 \to S$ that, given a finite sequence of states 
(representing the history of the play so far) which ends in a player~1 
state, chooses the next state.
The strategy must choose only available successors, i.e., for all $w \in S^*$ 
and $s \in S_1$ we have $\straa(w \cdot s) \in E(s)$.
The strategies for player~2 are defined analogously.
We write $\Straa$ and $\Strab$ for the sets of all strategies for 
player~1 and player~2, respectively.
An important special class of strategies are \emph{memoryless} strategies.
The memoryless strategies do not depend on the history of a play, 
but only on the current state. 
Each memoryless strategy for player~1 can be specified as a function
$\straa$: $S_1 \to S$ such that $\straa(s) \in E(s)$ for all $s \in S_1$,
and analogously for memoryless player~2 strategies.
Given a starting state $s\in S$, a strategy $\straa\in\Straa$ for player~1, 
and a strategy $\strab\in\Strab$ for player~2, there is a unique play, 
denoted $\pat(s,\straa,\strab)=\sseq$, which is defined as follows: 
$s_0=s$ and for all $k \geq 0$,
if $s_k \in S_1$, then $\straa(s_0,s_1,\ldots,s_k)=s_{k+1}$, and
if $s_k \in S_2$, then $\strab(s_0,s_1,\ldots,s_k)=s_{k+1}$.

\smallskip\noindent{\em B\"uchi and coB\"uchi objectives.}
We consider game graphs
with a B\"uchi objective for player~1 and the complementary 
coB\"uchi objective for player~2.
For a play $\pat = \sseq\in \Omega$,   
we define $\Inf(\pat) = 
\set{s \in S \mid \mbox{$s_k = s$ for infinitely many $k \geq 0$}}$
to be the set of states that occur infinitely often in~$\pat$.
We also define reachability and safety objectives as they will be useful in the
analysis of the algorithms.
\begin{enumerate}

 \item 
  \emph{Reachability and safety objectives.}
  Given a set $T \subseteq S$ of states, the reachability objective 
  $\Reach{T}$ requires that some state in $T$ be visited,
  and dually, 
  the safety objective $\Safe{F}$ requires that only states in $F$ 
  be visited.
  Formally, the sets of winning plays are
  $\Reach{T}= \set{\sseq \in \Pat \mid 
  \exists k \geq 0. \ s_k \in T}$
  and 
  $\Safe{F}=\set{\sseq \in \Pat \mid 
  \forall k \geq 0.\ s_k \in F}$.
  The reachability and safety objectives are dual in the sense that 
  $\Reach{T}= \Pat \setminus \Safe{S \setminus T}$.

\item
  \emph{B\"uchi and co-B\"uchi objectives.}
  Given a set $B \subseteq S$ of states, the B\"uchi objective 
  $\Buchi{B}$ requires that some state in $B$ be visited
  infinitely often, and dually, 
  the co-B\"uchi objective $\coBuchi{C}$ requires that only 
  states in $C$ be visited infinitely often.
  Thus, the sets of winning plays are
  $\Buchi{B}= \set{\pat \in \Pat \mid 
  \Inf(\pat)\cap B\neq\emptyset}$
  and 
  $\coBuchi{C}=\set{\pat \in \Pat \mid 
  \Inf(\pat)\subseteq C}$.
  The B\"uchi and coB\"uchi objectives are dual in the sense that 
  $\Buchi{B}= \Pat \setminus \coBuchi{S\setminus B}$.

\end{enumerate}

\smallskip\noindent{\em Winning strategies and sets.}
Given an objective $\Phi\subseteq\Pat$ for player~1, a strategy 
$\straa\in\Straa$ is a \emph{winning strategy}
for player~1 from a state $s$ if for all player~2 strategies $\strab\in\Strab$ 
the play $\pat(s,\straa,\strab)$ is winning, i.e., 
$\pat(s,\straa,\strab) \in \Phi$.
The winning strategies for player~2 are defined analogously.
A state $s\in S$ is winning for player~1 with respect to the objective 
$\Phi$ if player~1 has a winning strategy from $s$.
Formally, the set of \emph{winning states} for player~1 with respect to 
the objective $\Phi$ is 
$W_1(\Phi) =\set{s \in S \mid \exists \straa\in\Straa. 
\ \forall \strab\in\Strab.\ \pat(s,\straa,\strab) \in \Phi}.$
Analogously, the set of winning states for player~2 with respect to an 
objective $\Psi\subseteq\Pat$ is 
$W_2(\Psi) =\set{s \in S \mid \exists \strab\in\Strab. \ 
\forall \straa\in\Straa.\ \pat(s,\straa,\strab) \in \Psi}.$
We say that there exists a memoryless winning strategy 
for player~1 with respect to the objective $\Phi$ if there exists such 
a strategy from all states in $W_1(\Phi)$; and similarly for player~2.

\begin{theorem}[Classical memoryless determinacy]
\label{thrm:determinacy}
The following assertions hold.
\begin{enumerate}
\item For all game graphs $G=((S,E),(S_1,S_2))$, all B\"uchi objectives $\Phi$ 
for player~1, and the complementary coB\"uchi objective 
$\Psi=\Pat \setminus \Phi$ for player~2,
we have $W_1(\Phi)=S \setminus W_2(\Psi)$.

\item For all game graphs and all 
B\"uchi objectives $\Phi$ for player~1 and the complementary coB\"uchi 
objective $\Psi$ for player~2, there exists a memoryless winning 
strategy for both players.

\end{enumerate}
\end{theorem}

Observe that for B\"uchi objective $\Phi$ and the coB\"uchi 
objective $\Psi= \Pat \setminus \Phi$ by definition we have 
$S \setminus W_2(\Psi)= \set{s \in S \mid \forall \strab \in \Strab. \ \exists
\straa \in \Straa. \ \pat(s,\straa,\strab) \in \Phi }$.
Theorem~\ref{thrm:determinacy} states that 
$S\setminus W_2(\Psi)= \set{s \in S \mid \exists \straa \in \Straa. \ \forall 
\strab \in \Strab. \ \pat(s,\straa,\strab) \in \Phi}$, i.e., the order of
the universal and the existential quantifiers can be exchanged. 

\section{Iterative Algorithms for B\"uchi Games}
In this section we present the classical iterative algorithm for B\"uchi games
and an alternative iterative algorithm, 
i.e., algorithms to compute the winning sets in B\"uchi games.
The running time of the alternative algorithm is never more than the 
running time the classical algorithm by an additive factor of $O(m)$, where 
$m$ is the number of edges in the game graph.
We also present a family of game graphs with B\"uchi objectives where the
classical algorithm requires quadratic time and the alternative 
algorithm works in linear time.
We start with the notion of \emph{closed sets} and \emph{attractors} which
are key notions for the analysis of the algorithm.

\smallskip\noindent{\em Closed sets.} 
A set $U\subseteq S$ of states is a \emph{closed set} 
for player~1 if the following two conditions hold:
(a)~for all states $u \in (U \cap S_1)$, we have $E(u) \subseteq U$, 
i.e., all successors of player~1 states in $U$ are again in $U$; and
(b)~for all $u \in (U \cap S_2)$, we have 
$E(u) \cap U \neq \emptyset$, i.e., 
every player~2 state in $U$ has a successor in $U$.
The closed sets for player~2 are defined analogously.
Every closed set $U$ for player~$\ell$, for $\ell\in\set{1,2}$, 
induces a sub-game graph, denoted $G \restr U$.

\begin{proposition}\label{prop:closed}
Consider a game graph $G$, and a closed set $U$ for player~1.
Then the following assertions hold.
\begin{enumerate}
\item There exists a memoryless strategy $\strab$ for player~2 such that 
for all strategies $\straa$ for player~1 and for all states $s \in U$ we
have $\pat(s,\straa,\strab) \in \Safe{U}$.

\item For all $T \subseteq S \setminus U$, we have $W_1(\Reach{T}) \cap U
=\emptyset$.
\end{enumerate}
\end{proposition}

\noindent{\em Attractors.} 
Given a game graph $G$, a set $U \subseteq S$ of states, 
and a player $\ell\in\set{1,2}$, 
the set $\Attr_\ell(U,G)$ contains the states from which player~$\ell$ 
has a strategy to reach a state in $U$ against all strategies
of the other player;
that is, $\Attr_\ell(U,G) = W_\ell(\Reach{U})$.
The set $\Attr_1(U,G)$ can be computed inductively as follows:
let $R_0=U$; let 
\[
R_{i+1}= R_i \cup \set{s \in S_1 \mid E(s) \cap R_i \neq \emptyset} 
    \cup \set{s \in S_2 \mid E(s) \subseteq R_i}
\qquad \text{for all } i\ge 0;
\]
then $\Attr_1(U,G)= \bigcup_{i\ge 0} R_i$.
The inductive computation of $\Attr_2(U,G)$ is analogous.
For all states $s \in \Attr_1(U,G)$, define 
$\rank(s)=i$ if $s \in R_i \setminus R_{i-1}$,
that is, $\rank(s)$ denotes the least $i\ge 0$ such that $s$ is 
included in $R_i$.
Define a memoryless strategy $\straa\in\Straa$ for player~1 as follows: 
for each state $s \in (\Attr_1(U,G) \cap S_1)$ with $\rank(s)=i$, 
choose a successor $\straa(s)\in (R_{i-1} \cap E(s))$ 
(such a successor exists by the inductive definition).
It follows that for all states $s \in \Attr_1(U,G)$ and all strategies 
$\strab\in\Strab$ for player~2, the play $\pat(s,\straa,\strab)$ reaches 
$U$ in at most $|\Attr_1(U,G)|$ transitions.

\begin{proposition}\label{prop:attractor}
For all game graphs $G$, all players $\ell\in\set{1,2}$, and 
all sets $U\subseteq S$ of states, 
the set $S\setminus \Attr_\ell(U,G)$ is a closed set for player~$\ell$.
\end{proposition}

\subsection{Classical algorithm for B\"uchi games}
In this subsection we present the classical algorithm for B\"uchi games.
We start with an informal description of the algorithm.

\medskip\noindent{\bf Informal description of classical algorithm.}
The \emph{classical algorithm} (Algorithm~\ref{algorithm:classical})
works as follows.  
We describe an iteration $i$ of the algorithm: the set of states at iteration
$i$ is denoted as $S^i$, the game graph as $G_i$ and the set of
B\"uchi states $B\cap S^i$ as $B_i$.
At iteration $i$, the algorithm first finds the set of states $R_i$ from which 
player~1 has a strategy to reach the  set $B_i$, i.e., computes
$\Attr_1(B_i,G_i)$.
The rest of the states $\Tr_i = S^i \setminus R_i$ 
is a closed subset for player~1, and $\Tr_i \cap B_i =\emptyset$.
The set $\Tr_i$ is identified as winning for player~2.
Then the set of states $W_{i+1}$, from which player~2 has a strategy 
to reach the set $\Tr_i$, i.e., $\Attr_2(\Tr_i,G_i)$ is computed. 
The set $W_{i+1}$ is identified as a subset of the winning set for player~2
and it is removed from the state set.  
The algorithm then iterates on the reduced game graph.
Observe that at every iteration the set of states removed is an attractor set
and by Proposition~\ref{prop:attractor} the reduced game graph
(the complement of an attractor) is a closed set and hence a game graph. 
In every iteration it performs a \emph{backward} search from the
current B\"uchi states to find the set of states which can reach the B\"uchi 
set.
Each iteration takes $O(m)$ time and the algorithm runs for at most $O(n)$
iterations, where $m$ and $n$ denote the number of edges and states in the
game graph, respectively.
The algorithm is formally described as Algorithm~\ref{algorithm:classical}.
The correctness of the algorithm easily follows from the results in
\cite{McN93,Thomas97}.

\begin{algorithm}[t]
\caption{\bf Classical algorithm for B\"uchi Games}
\label{algorithm:classical}
{ 
\begin{tabbing}
aa \= aa \= aaa \= aaa \= aaa \= aaa \= aaa \= aaa \kill
\> {\bf Input :} A 2-player game graph $G=((S,E),(S_1,S_2)$ and 
 $B \subseteq S$. \\ 
\> {\bf Output:} $W\subseteq S$. \\
\> 1. $G_0 := G$; $S^0:= S$;   2. $W_0 := \emptyset$;  3. $i := 0$\\ 
\> 4. {\bf repeat }   \\ 
\>\>\> 4.1 $W_{i+1} := \avoidset(G_i,B \cap S^i)$ \\ 
\>\>\> 4.2 $S^{i+1} := S^i \setminus W_{i+1}$; $G_{i+1} = G \restr S^{i+1}$;  
4.3 $i := i+1$; \\
\>\> 
  {\bf until} $W_i = \emptyset$ \\
\> 5. {\bf return} $ W := \bigcup_{k=1}^{i} W_k$. \\

{\bf Procedure} $\avoidset$ \\ 
\> {\bf Input:} Game graph  $G_i$ and $B_i \subseteq S^i$. 
\\ \> 
{\bf Output:} set $W_{i+1} \subseteq S^i$. \\
\> 1. $R_i := \Attr_1(B_i, G_i)$;  2. $\Tr_i := S^i \setminus R_i$; 
3. $W_{i+1} := \Attr_{2}(\Tr_i, G_i)$ 
\end{tabbing}
}
\end{algorithm}

\begin{theorem}[Correctness and running time]\label{thrm:classical}
Given a game graph $G=((S,E),(S_1,S_2))$ and $B \subseteq S$ the following 
assertions hold:
\begin{enumerate}
\item $W=W_2(\coBuchi{S\setminus B})$ and $S\setminus W= W_1(\Buchi{B})$,
	where $W$ is the output of Algorithm~\ref{algorithm:classical}; 
\item the running time of Algorithm~\ref{algorithm:classical} is 
	$O(b\cdot m)$ where $b=|B|$ and $m=|E|$.
\end{enumerate}  
\end{theorem}

\begin{remark}
Observe that the size of the set of B\"uchi states $B$ can be $O(n)$, 
where $n=|S|$ is the number of states, i.e., $b$ in Theorem~\ref{thrm:classical}
can be $O(n)$. 
Hence the worst case running time of the classical algorithm can be 
$O(n \cdot m)$, where $n=|S|$ and $m=|E|$.
\end{remark}

\subsection{Alternative algorithm for B\"uchi games}
We now present a new alternative iterative algorithm for B\"uchi games.
The algorithm differs from the classical algorithm in its computation of
the set $\Tr_i$ (computed in step 2 of procudure $\avoidset$) at every 
iteration.
Recall that the set $\Tr_i$ is a player~1 closed set with empty intersection 
with the set of B\"uchi states. 
The alternative algorithm at every iteration identifies the set $\Tr_i$ 
in an alternative way. 
We first present an informal description of the algorithm.

\medskip\noindent{\bf Informal description of alternative algorithm.}
We describe an iteration $i$ of Algorithm~\ref{algorithm:alternative}.
We denote by $C=S \setminus B$ the set of coB\"uchi states.
We denote the set of states at iteration $i$ by $S^i$, 
the game graph as $G_i$, the set of B\"uchi states $B\cap S^i$ as $B_i$, and
the set of coB\"uchi states as $C \cap S^i$ as $C^i$.
The algorithm proceeds as follows: first it computes the set of player~1 
states in $C^i$ with all successors in $C^i$ and the set of player~2 
states in $C^i$ with a successor in $C^i$. 
Then the player~2 attractor to the union of the above two sets is computed
and let this set be $Z_i$. 
The states of $Z_i$ such that player~1 has a strategy to leave $Z_i$ is not 
a part of the player~1 closed set, and the remaining states of $Z_i$ is 
a player~1 closed set with empty intersection with $B_i$, 
and this is identified as the set similar to the set $\Tr_i$ of 
Algorithm~\ref{algorithm:classical}.
The details of the algorithms is as follows.
In step~4.2 the set of player~1 states $C_1^i$ is computed where for all 
$s \in C_1^i$, all successors of $s$ in $G_i$ is in $C^i$;
and in step~4.3 the set of player~2 states $C_2^i$ is computed where for all 
$s \in C_2^i$, there is a successor of $s$ in $C^i$.
Then the set $X_i$ is computed as the set of states such that player~2 
has a strategy in $G_i$ to reach $C_1^i \cup C_2^i$ against all player~1 
strategies, i.e., $X_i = \Attr_2(C_1^i \cup C_2^i,G_i)$, and 
the set $Z_i$ is obtained as $Z_i=X_i \cap C^i$.
The set $D_i$ denotes the set of states such that player~1 can escape $Z_i$
in one step, i.e., either a player~1 state with an edge out of $Z_i$ or a 
player~2 state with all edges out of $Z_i$.
The set $L_i$ denotes the set of states where player~1 has a strategy in 
$X_i$ to reach $D_i$ against all player~2 strategies, i.e.,
$L_i =\Attr_1(D_i, G_i \restr X_i)$.
Observe that the set $X_i$ is not always a proper sub-game, however, for 
all states in $s \in X_i \setminus D_i$ we have $E(s) \cap X_i
\neq \emptyset$, and hence for the purpose of the computation of the 
attractor of $D_i$ we can consider $G_i \restr X_i$ as a sub-game.
The set $\wh{\Tr}_i = Z_i \setminus L_i$ is identified as winning for player~2.
Then the set of states $W_{i+1}$, from which player~2 has a strategy 
to reach the set $\wh{\Tr}_i$, i.e., $\Attr_2(\wh{Tr}_i,G_i)$ is computed. 
The set $W_{i+1}$ is identified as a subset of the winning set for player~2
and it is removed from the state set.  
The algorithm then iterates on the reduced game graph. 
The algorithm is described formally as Algorithm~\ref{algorithm:alternative}.

\begin{algorithm}[t]
\caption{\bf Alternative algorithm for B\"uchi Games}
\label{algorithm:alternative}
{ 
\begin{tabbing}
aa \= aa \= aaa \= aaa \= aaa \= aaa \= aaa \= aaa \kill
\> {\bf Input :} A 2-player game graph $G=((S,E),(S_1,S_2)$ and 
 $B \subseteq S$. 
\\ \> 
{\bf Output:} $W\subseteq S$. \\
\> 1. $G_0 := G$; $S^0:= S$;  $C = S \setminus B$;  2. $W_0 := \emptyset$;  
3. $i := 0$\\ 
\> 4. {\bf repeat }   \\ 
\>\> 4.1 $C^i := C \cap S^i$;\\
\>\> 4.2 $C_1^i := \set{s \in S_1\cap C^i\mid E(s)\cap S^i \subseteq C^i}$;\\
\>\> 4.3 $C_2^i := \set{s \in S_2\cap C^i\mid E(s)\cap C^i \neq \emptyset}$;\\
\>\> 4.4 $X_i := \Attr_2(C_1^i \cup C_2^i, G_i)$; \\
\>\> 4.5 $Z_i := X_i \cap C^i$; \\
\>\> 4.6 $D_i := 
\set{s\in S_1\cap Z_i\mid E(s)\cap S^i\cap(S^i\setminus Z_i) \neq \emptyset}$ 
\\ \>\>\>\>  
$\cup 
\set{s\in S_2 \cap Z_i\mid E(s) \cap S^i \subseteq (S^i \setminus Z_i)} \cup
(X_i \setminus Z_i)$. \\ 
 
\>\> 4.7 $L_i= \Attr_1(D_i, G_i \restr X_i)$; \\
\>\> 4.8 $\wh{\Tr}_i =Z_i \setminus L_i$; \\
\>\> 4.9 $W_{i+1} := \Attr_2(\wh{\Tr}_i,G_i)$;\\ 
\>\> 4.10 $S^{i+1} := S^i \setminus W_{i+1}$; $G_{i+1} = G \restr S^{i+1}$; $i := i+1$; \\
\> 
  {\bf until} $W_i = \emptyset$ \\
\> 5. {\bf return } $ W := \bigcup_{k=1}^{i} W_k$. 
\end{tabbing}
}
\end{algorithm}


\medskip\noindent{\bf Correctness arguments.} 
The main argument to prove the correctness of 
Algorithm~\ref{algorithm:alternative} is as follows: we will show that,
given that the game graph $G_i$ are same at iteration $i$ of 
Algorithm~\ref{algorithm:classical} and Algorithm~\ref{algorithm:alternative}, 
set $\Tr_i$ computed in step~2 of the iteration of classical algorithm and
the set $\wh{\Tr}_i$ computed in step~4.8 of the alternative algorithm coincide.
Once we prove this result the correctness of the alternative algorithm follows
easily.
We prove this result in several steps.
The following proposition states that $\Tr_i$ is the largest player~1 
closed subset of $C^i$ and it follows easily from the properties of 
attractors and Proposition~\ref{prop:closed}.

\begin{proposition}\label{prop:alt1}
Let $G_i$ be the graph at iteration $i$ if Algorithm~\ref{algorithm:classical} 
and let $U_i \subseteq C^i$ such that $U_i$ is player~1 closed,
then $U_i \subseteq \Tr_i$.
\end{proposition}

By Proposition~\ref{prop:alt1} to prove our desired claim it suffices to 
show that $\Tr_i \subseteq \wh{\Tr}_i$, and $\wh{\Tr}_i$ is a player~1 closed
subset of $C^i$ (this would imply $\wh{\Tr}_i \subseteq \Tr_i$).

\begin{lemma}\label{lemm:alt1}
$\Tr_i \subseteq C_1^i \cup C_2^i$.
\end{lemma}
\begin{proof}
Observe that $\Tr_i$ is a player~1 closed of $C^i$.
Hence for all states $s \in \Tr_i$ the following assertions hold:
(a)~if $s \in S_1$, then $E(s) \cap S^i \subseteq \Tr_i \subseteq C^i$;
and 
(b)~if $s \in S_2$, then 
$E(s) \cap \Tr_i \neq \emptyset$, and hence 
$E(s) \cap C^i \neq \emptyset$. 
Hence if $s \in \Tr_i$, then $s \in C_1^i \cup C_2^i$.
\hfill\qed
\end{proof}

\begin{lemma}\label{lemm:alt2}
$\Tr_i \subseteq Z_i$, where $Z_i$ is the set computed at step~4.5 of 
Algorithm~\ref{algorithm:alternative}.
\end{lemma}
\begin{proof}
By Lemma~\ref{lemm:alt1} we have $\Tr_i \subseteq C_1^i \cup C_2^i$
and hence we have 
$\Tr_i \subseteq \Attr_2(C_1^i \cup C_2^i, G_i)$.
Since $\Tr_i \subseteq C^i$, we have $\Tr_i \subseteq X_i \cap C^i =Z_i$.
\hfill\qed
\end{proof}

\begin{lemma}\label{lemm:alt3}
$\Tr_i \subseteq Z_i \setminus L_i$, where $Z_i$ and $L_i$ are the sets 
computed at step~4.5 and 4.7 of Algorithm~\ref{algorithm:alternative},
respectively.
\end{lemma}
\begin{proof}
By Lemma~\ref{lemm:alt2} we have $\Tr_i \subseteq Z_i$ and hence we have 
$S^i \setminus Z_i \subseteq S^i \setminus \Tr_i$.
Observe that from states in $D_i$ player~1 can force the game to reach 
$(S^i \setminus Z_i)$.
Similarly, in the sub-game $X_i$ with $D_i$ as target set, player~1 can force
the game to reach $D_i$ from $L_i$ and then force the game to reach 
$S^i \setminus Z_i$.
Since $\Tr_i \subseteq Z_i \subseteq X_i$ and $\Tr_i$ is a player~1 closed set,
player~2 can keep the game in $\Tr_i$ forever 
(by Proposition~\ref{prop:closed}).
Hence we must have $\Tr_i \cap L_i=\emptyset$.
Hence we have $\Tr_i \subseteq Z_i$ and $\Tr_i \cap L_i =\emptyset$.
Thus we obtain $\Tr_i \subseteq Z_i \setminus L_i$.
\hfill\qed
\end{proof}

\begin{lemma}\label{lemm:alt4}
$Z_i \setminus L_i \subseteq C^i$ and $Z_i \setminus L_i$ is a player~1 
closed set.
\end{lemma}
\begin{proof}
Since $X_i =\Attr_2(C_1^i \cup C_2^i,G_i)$, it follows that for all 
states $s \in X_i$ we have 
(a) if $s \in S_1 \cap X_i$, then $E(s) \cap S^i \subseteq X_i$, and
(b) if $s \in S_2 \cap X_i$, then we have $E(s) \cap X_i \neq \emptyset$.
Since $L_i=\Attr_1(D_i, G_i \restr X_i)$, for all states 
$s \in (Z_i \setminus L_i)$ the following assertions hold:
(a)~if $s \in S_1$, then $E(s) \cap S^i \subseteq (Z_i \setminus L_i)$,
	as otherwise $s$ would have been in $L_i$; and
(b)~if $s \in S_2$, then $E(s) \cap (Z_i \setminus L_i) \neq \emptyset$, 
	as otherwise $s$ would have been in $L_i$.
Hence it follows that $(Z_i\setminus L_i)$ is player~1 closed, and
since $Z_i \subseteq C^i$, the desired result follows. 
\hfill\qed
\end{proof}

\begin{lemma}\label{lemm:alt5}
$\Tr_i =\wh{\Tr}_i$.
\end{lemma}
\begin{proof}
By Lemma~\ref{lemm:alt3} we have $\Tr_i \subseteq \wh{\Tr}_i$.
By Lemma~\ref{lemm:alt4} we have $\wh{\Tr}_i$ is a player~1 closed subset of
$C^i$.
Then, by Proposition~\ref{prop:alt1} we have $\wh{\Tr}_i \subseteq \Tr_i$.
This proves the desired result.
\hfill\qed
\end{proof}

The correctness of Algorithm~\ref{algorithm:alternative} easily follows from
Lemma~\ref{lemm:alt5} and induction.

\begin{theorem}[Correctness of Algorithm~\ref{algorithm:alternative}]
\label{thrm:alt-correctness}
Given a game graph $G=((S,E),(S_1,S_2))$ and $B \subseteq S$ we have  
$W=W_2(\coBuchi{S\setminus B})$ and $S\setminus W= W_1(\Buchi{B})$,
where $W$ is the output of Algorithm~\ref{algorithm:alternative}.
\end{theorem}

\medskip\noindent{\bf Work analysis of Algorithm~\ref{algorithm:alternative}.}
We first analyze the work for the computation of step~4.2 and step~4.3 over 
all iterations of Algorithm~\ref{algorithm:alternative}.
It easily follows that $C_2^i = C_2^0 \cap S^i$, this follows since if a state
$s \in S_2$ has an edge to $W_{i-1}$, then $s$ would have been in 
$W_{i-1}$ itself.
The total work in the computation of the sets $C_1^i$ overall iterations is
$O(m)$: this is achieved as follows.
For every state $s \in S_1$ we keep a counter for the number of edges to 
the set $S \setminus C=B$, and once an edge to $S\setminus C$ 
is removed from the graph the counter for the respective state is decremented 
by~1.
Once the counter for a state reaches~0 it is included in $C_1^i$.
Hence the total work for step~4.2 and step~4.3 overall iterations is $O(m)$.
We now argue that the excess work of Algorithm~\ref{algorithm:alternative}
as compared to the classical algorithm is at most $O(m)$.
The total work of step~4.2 and step~4.3 is already bounded by $O(m)$.
The rest of the argument is as follows: the classical algorithm for the 
computation of $\Tr_i$ never works on the edges in $\Tr_i$, i.e., on
edges in $E \cap (\Tr_i \times \Tr_i)$.
At iteration~$i$ Algorithm~\ref{algorithm:alternative} does excess work as 
compared to classical algorithm on the edges in $\Tr_i$ and does only 
constant amount of work on this edges.
However, edges in $\Tr_i$ are removed at every iteration $i$, and hence
the excess work of Algorithm~\ref{algorithm:alternative} as compared to
the classical algorithm is $O(m)$.

\begin{theorem}[Running time of Algorithm~\ref{algorithm:alternative}]
\label{thrm:run-alternative}
Let $\RT_1(G,B)$ denote the running time of Algorithm~\ref{algorithm:classical}
on a game graph $G$ with $B\subseteq S$ and $\RT_2(G,B)$ denote the 
running time of Algorithm~\ref{algorithm:alternative}.
Then we have $\RT_2(G,B) \leq \RT_1(G,B) + O(m)$, where $m=|E|$.
\end{theorem}

We now present an example of a  family of game graphs where the classical 
algorithm requires quadratic time, whereas 
Algorithm~\ref{algorithm:alternative} works in linear time.

\begin{example}
The family of game graphs is constructed as follows.
Given $n\geq 1$ we consider a game graph consisting of $n$ gadgets 
${\cal H}(0), {\cal H}(1), \ldots {\cal H}(n)$ as follows.
The gadget ${\cal H}(i)$ consits of a player~1 state $t_i$ (shown as a 
$\bigcirc$-state in Fig~\ref{figure:buchi-example}) and a player~2 state 
$w_i$ (shown as a $\Diamond$ state in Fig~\ref{figure:buchi-example}).
The set of edges is as follows:
\begin{enumerate}
\item For $0 \leq i < n$ we have $E(w_i)= \set{t_i,t_{i+1}}$ and 
	$E(w_n)=\set{t_n}$.

\item For $0 < i \leq n$ we have $E(t_i)=\set{t_i,w_{i-1}}$ and 
	$E(t_0)=\set{t_0}$.
\end{enumerate}
The set of B\"uchi states is $B=\set{w_i \mid 0 \leq i \leq n}$.
Given $n\geq 1$, the game graph constructed has $2n$ states and $4n-2$ edges, i.e., 
we have $O(n)$ states and $O(n)$ edges.
\begin{figure}[t]
\begin{center}
\setlength{\unitlength}{0.00033333in}
\begingroup\makeatletter\ifx\SetFigFont\undefined%
\gdef\SetFigFont#1#2#3#4#5{%
  \reset@font\fontsize{#1}{#2pt}%
  \fontfamily{#3}\fontseries{#4}\fontshape{#5}%
  \selectfont}%
\fi\endgroup%
{\renewcommand{\dashlinestretch}{30}
\begin{picture}(11144,3352)(0,-10)
\put(4972,2640){\makebox(0,0)[lb]{{\SetFigFont{8}{9.6}{\rmdefault}{\mddefault}{\updefault}$w_2$}}}
\thicklines
\put(8459.500,1215.000){\arc{604.669}{5.7640}{9.9439}}
\blacken\path(8806.277,1132.412)(8722.000,1365.000)(8686.909,1120.115)(8806.277,1132.412)
\put(6284.500,1215.000){\arc{604.669}{5.7640}{9.9439}}
\blacken\path(6631.277,1132.412)(6547.000,1365.000)(6511.909,1120.115)(6631.277,1132.412)
\put(1634.500,1215.000){\arc{604.669}{5.7640}{9.9439}}
\blacken\path(1981.277,1132.412)(1897.000,1365.000)(1861.909,1120.115)(1981.277,1132.412)
\put(10522,1515){\ellipse{540}{540}}
\put(8422,1515){\ellipse{540}{540}}
\put(6247,1515){\ellipse{540}{540}}
\put(1597,1515){\ellipse{540}{540}}
\path(9584,3053)(9284,2753)(9584,2453)
	(9884,2753)(9584,3053)
\path(7409,3053)(7109,2753)(7409,2453)
	(7709,2753)(7409,3053)
\path(5122,3090)(4822,2790)(5122,2490)
	(5422,2790)(5122,3090)
\path(697,3090)(397,2790)(697,2490)
	(997,2790)(697,3090)
\path(9097,3315)(11122,3315)(11122,615)
	(9097,615)(9097,3315)
\path(6847,3315)(8872,3315)(8872,615)
	(6847,615)(6847,3315)
\path(4672,3315)(6697,3315)(6697,615)
	(4672,615)(4672,3315)
\path(22,3315)(2047,3315)(2047,615)
	(22,615)(22,3315)
\path(9697,2565)(10372,1740)
\blacken\path(10173.585,1887.755)(10372.000,1740.000)(10266.460,1963.744)(10173.585,1887.755)
\path(9397,2640)(8497,1740)
\blacken\path(8624.279,1952.132)(8497.000,1740.000)(8709.132,1867.279)(8624.279,1952.132)
\path(8647,1665)(9472,2490)
\blacken\path(9344.721,2277.868)(9472.000,2490.000)(9259.868,2362.721)(9344.721,2277.868)
\path(7555,2592)(8230,1767)
\blacken\path(8031.585,1914.755)(8230.000,1767.000)(8124.460,1990.744)(8031.585,1914.755)
\path(7184,2678)(6284,1778)
\blacken\path(6411.279,1990.132)(6284.000,1778.000)(6496.132,1905.279)(6411.279,1990.132)
\path(6509,1703)(7334,2528)
\blacken\path(7206.721,2315.868)(7334.000,2528.000)(7121.868,2400.721)(7206.721,2315.868)
\path(5313,2598)(5988,1773)
\blacken\path(5789.585,1920.755)(5988.000,1773.000)(5882.460,1996.744)(5789.585,1920.755)
\path(4972,2565)(4072,1665)
\blacken\path(4199.279,1877.132)(4072.000,1665.000)(4284.132,1792.279)(4199.279,1877.132)
\path(805,2591)(1480,1766)
\blacken\path(1281.585,1913.755)(1480.000,1766.000)(1374.460,1989.744)(1281.585,1913.755)
\path(1672,1740)(2497,2565)
\blacken\path(2369.721,2352.868)(2497.000,2565.000)(2284.868,2437.721)(2369.721,2352.868)
\dashline{90.000}(2497,1665)(3922,1665)
\put(10372,1440){\makebox(0,0)[lb]{{\SetFigFont{8}{9.6}{\rmdefault}{\mddefault}{\updefault}$t_0$}}}
\put(8272,1440){\makebox(0,0)[lb]{{\SetFigFont{8}{9.6}{\rmdefault}{\mddefault}{\updefault}$t_1$}}}
\put(6097,1440){\makebox(0,0)[lb]{{\SetFigFont{8}{9.6}{\rmdefault}{\mddefault}{\updefault}$t_2$}}}
\put(1447,1440){\makebox(0,0)[lb]{{\SetFigFont{8}{9.6}{\rmdefault}{\mddefault}{\updefault}$t_n$}}}
\put(7222,2640){\makebox(0,0)[lb]{{\SetFigFont{8}{9.6}{\rmdefault}{\mddefault}{\updefault}$w_1$}}}
\put(9397,2640){\makebox(0,0)[lb]{{\SetFigFont{8}{9.6}{\rmdefault}{\mddefault}{\updefault}$w_0$}}}
\put(247,90){\makebox(0,0)[lb]{{\SetFigFont{8}{9.6}{\rmdefault}{\mddefault}{\updefault}${\cal H}(n)$}}}
\put(4972,90){\makebox(0,0)[lb]{{\SetFigFont{8}{9.6}{\rmdefault}{\mddefault}{\updefault}${\cal H}(2)$}}}
\put(7222,90){\makebox(0,0)[lb]{{\SetFigFont{8}{9.6}{\rmdefault}{\mddefault}{\updefault}${\cal H}(1)$}}}
\put(9247,90){\makebox(0,0)[lb]{{\SetFigFont{8}{9.6}{\rmdefault}{\mddefault}{\updefault}${\cal H}(0)$}}}
\put(547,2640){\makebox(0,0)[lb]{{\SetFigFont{8}{9.6}{\rmdefault}{\mddefault}{\updefault}$w_n$}}}
\put(10559.500,1215.000){\arc{604.669}{5.7640}{9.9439}}
\blacken\path(10906.277,1132.412)(10822.000,1365.000)(10786.909,1120.115)(10906.277,1132.412)
\end{picture}
}  
\end{center}
\caption{}\label{figure:buchi-example}
\end{figure}

 The sets $\Tr_i$ and $W_{i+1}$ obtained at iteration $i$, for $i\geq 0$, for 
the classical algorithm  are as follows: 
$\Tr_{i}=\set{t_{i}}$ and $W_{i+1}=\set{t_{i},w_{i}}$.
At iteration $i$ the classical algorithm works on edges of gadgets 
${\cal H}(\ell)$, for $\ell \geq i+1$.
Hence the total work of the classical algorithm is at least 
$
\sum_{i=1}^{n-1} (n-i)= O(n^2),
$
i.e., the classical algorithm requires time quadratic in the size
of the game graph.

We now analyze the work of Algorithm~\ref{algorithm:alternative}.
The sets of Algorithm~\ref{algorithm:alternative} of iteration $i$, 
for $i \geq 0$, is as follows: 
(a)~$C_1^i=\set{t_{i}}$ and $C_2^i=\emptyset$;
(b)~$X_i=\set{t_{i},w_{i}}$ and $Z_i=\set{t_{i}}$; and 
(c)~$D_i=\set{t_{i}}$, $L_i=\set{w_{i}}$ and $\wh{\Tr}_i=\set{t_{i}}$.
The work of Algorithm~\ref{algorithm:alternative} is constant for every 
iteration $i$: since the computation of steps 4.4 to steps 4.9 only works 
on edges in ${\cal H}(i)$ and ${\cal H}(i+1)$.
Hence the total work of Algorithm~\ref{algorithm:alternative} is 
$O(n)$, i.e., the alternative algorithm works in linear time.

 Also if in every gadget ${\cal H}(i)$ the self-loop at state $t_i$ is
replaced by a cycle of $2\log(n)$ states, then both the classical 
algorithm and the algorithm of~\cite{CJH03} requires $O(n^2 \log (n))$ time,
whereas Algorithm~\ref{algorithm:alternative} requires $O(n \log(n))$ time. 
\hfill\qed
\end{example}

\medskip\noindent{\bf Dovetailing Algorithm~\ref{algorithm:classical} and
Algorithm~\ref{algorithm:alternative}.}
We already proved that the set $\Tr_i$ and $\wh{\Tr}_i$ of 
Algorithm~\ref{algorithm:classical} and Algorithm~\ref{algorithm:alternative} 
coincide.
Algorithm~\ref{algorithm:classical} never works on the edges in $\Tr_i$ and
is favorable when $\Tr_i$ is large and 
Algorithm~\ref{algorithm:alternative} is favorable when $\Tr_i$ is small.
Hence in every iteration both the algorithms can be run in a dovetailing 
fashion and obtaining $\Tr_i$ by the algorithm that completes first.
The computation of the sets $C_1^{i+1}$ and $C_2^{i+1}$ can be computed during 
the computation of the set $W_{i+1}$.

\section{Improved Algorithm for B\"uchi Games}

 In this section we present the improved algorithm for B\"uchi games.
The algorithm is a generalization of the improved algorithm of~\cite{CJH03}
to general game graphs, as compared to the algorithm of~\cite{CJH03} which 
works only for binary game graphs (game graphs with every state having 
out-degree at most 2). 
We will use the following notations in this section.

\medskip\noindent{\bf Notation.} 
For a set $U$ and a game graph $G$ we denote by 
$\Source(U,G) = 
\set{s \in S \mid E(s) \cap U \neq \emptyset}$
is the set of states with edges that enter $U$.
Given a game graph $G$ we denote by $\delta(G)=\max\set{ |E(s)| \mid s \in S}$
the maximum out-degree of a state in $G$.

\medskip\noindent{\bf Informal description of the new algorithm}
We observe that in step~1 of every iteration~$i$ of the classical algorithm 
an $O(m)$ \emph{backward} alternating search is performed to compute the 
set~$R_i$, where $m$ is the number of edges.
The key idea of our \emph{improved algorithm} 
(Algorithm~\ref{algorithm:improved}) is to perform a cheap \emph{forward} 
exploration of edges in order to discover subsets of the winning set for 
player~2. 
%
Let $U$ be the set of sources of edges entering the winning set of
player~2 discovered in the previous iteration.
The states in set $U$ are new candidates to be included in the
winning set of player~2.
The cheap forward exploration of edges is performed when the size of the
set $U$ is small.
Formally, if $|U| \geq \big(\frac{m}{\log(n)}\big)$, then an iteration of the 
classical algorithm is executed (step~4.1), i.e., the backward search is 
performed.
Otherwise, we perform the cheap forward search as follows:
we add an auxiliary state $\wh{s}$ with an edge to every state in $U$.  
From the state $\wh{s}$ a BFS is performed for  $\big(2\cdot\frac{m}{\log n}\big)$ 
steps in step~4.2.2 of Algorithm~\ref{algorithm:improved}.
In steps~4.2.3---4.2.7 we check if the explored subgraph contains a 
closed set for player~1 in which player~2 has a winning strategy.
If no such set is detected then one iteration of the classical
algorithm is executed.  
The key for an improved bound of our algorithm is the observation
that if step~4.2.7 fails to identify a non-empty winning subset for
player~2, then the set discovered by the following iteration of the
classical algorithm has at least $\big(\frac{\log n}{\log (\delta(G))}\big)$ 
states.
A formal presentation of the algorithm is given as Algorithm~\ref{algorithm:improved}.

\begin{algorithm}[t]
\caption{\bf Improved Algorithm for B\"uchi Games}
\label{algorithm:improved}
{ 
\begin{tabbing}
aa \= aa \= aaa \= aaa \= aaa \= aaa \= aaa \= aaa \= aaa \kill
\> {\bf Input :} A 2-player game graph $G=((S,E),(S_1,S_2)$ and 
 $B \subseteq S$. 
\\ \> 
{\bf Output:} $W\subseteq S$. \\
\> 1. $G_0 := G$; $S^0:= S$;  $C = S \setminus B$; 
2. $W_0 := \emptyset$;  3. $i := 0$\\ 
\> 4. {\bf repeat } \\
\>\> 4.1 {\bf if } ($|\Source(W_i,G)| \geq \frac{m}{\log(n)}$)\\
\>\>\> 4.1.1 $W_{i+1} := \avoidset(G_i, B \cap S^i)$ \\
\>\>\> 4.1.2 {\bf go to} step 4.3. \\
\>\> 4.2 {\bf else} \\
\>\>\> 4.2.1 {\bf add} a state  $\wh{s}$ and an edge fro $\wh{s}$ to a state in 
	$s \in \Source(W_i,G)$ \\
\>\>\> 4.2.2 Find the reachable subgraph $R_{\wh{s}}$ by a BFS for 
	$\big(2\cdot\frac{m}{\log n}\big)$ steps \\
\>\>\>  4.2.3 Let $F_{\wh{s}}$ denote the set of states in the frontier of the BFS \\
\>\>\> 4.2.4 $T_{\wh{s}} := \{ s \in S_2 \cap F_{\wh{s}} \mid E(s) \cap S^i \cap 
	R_{\wh{s}}  = \emptyset \} \cup (S_1 \cap F_{\wh{s}})$ \\
\>\>\> 4.2.5 $A_{\wh{s}} := 
\Attr_1((R_{\wh{s}}\cap B \cap S^i) \cup T_{\wh{s}},G_i \restr R_{\wh{s}})$ \\ 
\>\>\> 4.2.6 $\Tr_i := (R_{\wh{s}}\setminus A_{\wh{s}})$ \\
\>\>\> 4.2.7 {\bf if} $(\Tr_i \neq \emptyset)$ \\
\>\>\>\> 4.2.7.1 {\bf then} $W_{i+1} := \Attr_2(\Tr_i,G_i)$ \\   
\>\>\> 4.2.8 {\bf else} \\ 
\>\>\>\> 4.2.8.1  $W_{i+1} := \avoidset(G_i, B \cap S^i)$\\
\>\>\> 4.2.9 {\bf remove} the state $\wh{s}$;\\ \\

\>\> 4.3 $S^{i+1} := S^i \setminus W_{i+1}$; $G_{i+1} = G \restr S^{i+1}$; $i := i+1$; \\
\>\> 
  {\bf until} $W_i = \emptyset$ \\
\> 5. {\bf return } $ W := \bigcup_{k=1}^{i} W_k$.
\end{tabbing}
}
\end{algorithm}

\begin{theorem}[Correctness of Algorithm~\ref{algorithm:improved}]
\label{thrm:improved-correctness}
Given a game graph $G=((S,E),(S_1,S_2))$ and $B \subseteq S$ we have  
$W=W_2(\coBuchi{S\setminus B})$ and $S\setminus W= W_1(\Buchi{B})$,
where $W$ is the output of Algorithm~\ref{algorithm:improved}.
\end{theorem}
\begin{proof} 
We prove by induction that $W_i$ computed in any iteration of 
the improved algorithm satisfies $W_i \subseteq W_2(\coBuchi{S\setminus B})$.
  
\noindent{\em Base case:} 
$W_0 = \emptyset \subseteq W_2(\coBuchi{S\setminus B})$. 

\noindent{\em  Inductive case:} We argue that $W_i \subseteq 
W_2(\coBuchi{S\setminus B})$
implies that  $W_{i+1} \subseteq W_2(\coBuchi{S\setminus B})$. 
The case when step~4.1.1 gets executed, or step 4.2.7 fails and step 4.2.8 
gets executed, then the correctness follows from the correcntess of the 
iteration of the classical  algorithm.
We focus on the case when step~4.2.7 gets executed, i.e., a non-empty set
$\Tr_i$ is discovered as $(R_{\wh{s}} \setminus A_{\wh{s}})$.
For state $s \in S_1 \cap (R_{\wh{s}} \setminus F_{\wh{s}})$, 
we have $E(s) \cap S^i \subseteq R_{\wh{s}}$. 
It follows from step 4.2.4 that $F_{\wh{s}} \cap S_1 \subseteq T_{\wh{s}}$. 
Let $Z= (R_{\wh{s}} \cap B \cap S^i) \cup T_{\wh{s}}$.
Hence the following two conditions hold:
 \[ 
  (1).\ F_{\wh{s}} \cap S_1  \subseteq T_{\wh{s}} \subseteq 
  \Attr_1(Z,R_{\wh{s}})  \subseteq A_{\wh{s}}; \qquad 
  (2).\ R_{\wh{s}} \cap B \cap S^i \subseteq 
  \Attr_1(Z,R_{\wh{s}}) \subseteq A_{\wh{s}}.
\] 
By property of attractor we have the following property for 
$(R_{\wh{s}} \setminus A_{\wh{s}})$; 
for all states $s \in (R_{\wh{s}} \setminus A_{\wh{s}})$ the following 
assertions hold:
(a)~if $s \in S_1$, then $E(s) \cap S^i \subseteq 
(R_{\wh{s}} \setminus A_{\wh{s}})$, and
(b)~if $s \in S_2$, then $E(s) \cap S^i \cap 
(R_{\wh{s}} \setminus A_{\wh{s}}) \neq \emptyset$.

Hence  $(R_{\wh{s}} \setminus A_{\wh{s}})$ is a player~1 closed set in 
$G_i$ and $(R_{\wh{s}} \setminus A_{\wh{s}}) \cap B =\emptyset$.
It follows that 
$(R_{\wh{s}} \setminus A_{\wh{s}}) \subseteq W_2(\coBuchi{S \setminus B})$.
Hence it follows that 
$W_{i+1}\subseteq W_2(\coBuchi{S \setminus B})$.
The correctness of Algorithm~\ref{algorithm:improved} follows.
\hfill\qed
\end{proof}

\medskip\noindent{\bf Work analysis of algorithm~\ref{algorithm:improved}.}
We now focus on the work analysis of Algorithm~\ref{algorithm:improved}.
Let us denote by $\ell$ the depth of the search of the BFS at step~4.2.2 of 
Algorithm~\ref{algorithm:improved}.
Since $\Source(W_i,G) \leq \big(\frac{m}{\log(n)}\big)$ and 
the BFS proceeds for $\big(2\cdot\frac{m}{\log (n)}\big)$ steps, the BFS 
explores at least $\frac{m}{\log(n)}$ edges of the game graph $G_i$. 
Hence must have $\delta(G)^\ell \geq \big(\frac{m}{\log (n)}\big)$.
Thus we obtain that $\ell \geq O\big(\frac{\log (m)}{\log (\delta(G))}\big)
= O\big(\frac{\log (n)}{\log (\delta(G))}\big)$.
In the following lemma we denote by $\ell$ the depth of the BFS search 
at step~4.2.2 of Algorithm~\ref{algorithm:improved}.

\begin{lemma}\label{lemm:improved1}
Let $R_{\wh{s}}$ be the set computed in step 4.2.2 of 
Algorithm~\ref{algorithm:improved} and $\ell$ be the depth of the BFS in
step 4.2.2.
Let $W\subseteq R_{\wh{s}}$ be an player~1 closed set such that 
$W \cap B=\emptyset$ and $|W| < \ell$.
Then $W\subseteq (R_{\wh{s}} \setminus A_{\wh{s}})$, and hence
$W$ is discovered in step 4.2.7.
\end{lemma}
\begin{proof} Given a game graph $G$ and a set of states $U$ we define
sequences $U_i$ as follows:
\[
U_0  =  U; \qquad
U_{k+1}  =  U_k \cup \{ s \in S_1 \mid E(s) \cap S_k \neq \emptyset\}
          \cup  \{ s \in S_2 \mid E(s) \subseteq S_k \} \ \  k \geq 0. 
\]
By definition $\Attr_1(U,G)=\bigcup_k  U_k$.
We prove by induction that $W \cap A_{\wh{s}}=\emptyset$. 
By Step 4.2.5 we have 
$A_{\wh{s}}=\Attr_1((R_{\wh{s}} \cap B \cap S^i)\cup T_{\wh{s}},G_i \restr 
R_{\wh{s}})$. 
Let $U_i$ be the sequence of set of states 
in the attractor computation of 
$(R_{\wh{s}} \cap B \cap S^i)\cup T_{\wh{s}}$ 
with $U_0=(R_{\wh{s}} \cap B \cap S^i)\cup T_{\wh{s}}$. 
We show by induction that $U_i\cap W=\emptyset$.

\noindent{\em Base case.} 
Given $|W| \leq \ell-1$, for all states $w\in W$ there is a path from 
$\wh{s}$ of length at most $|W|\leq \ell-1$ to $w$.
It follows that in the BFS from $\wh{s}$ depth of any state $w\in W$ is 
less than $\ell$.
Hence we have $W \cap F_{\wh{s}}=\emptyset$. 
Since $T_{\wh{s}}\subseteq F_{\wh{s}}$  we have $W \cap T_{\wh{s}}=\emptyset$. 
Since $W \cap B=\emptyset$ we have $W \cap (B \cap S^i) =\emptyset$. 
It follows that 
$W\cap((R_{\wh{s}} \cap B \cap S^i)\cup T_{\wh{s}})=\emptyset$.
This proves the base case that $U_0\cap W=\emptyset$.

\noindent{\em Inductive case.}
Given $U_k\cap W=\emptyset$ we show that 
$U_{k+1} \cap W= \emptyset$. 
Since $W$ is a player~1 closed set, the following assertions hold:
for a state $s \in W \cap S_2$, we have $E(s) \cap W \neq \emptyset$ and
for a state $s \in W \cap S_1$, we have $E(s) \subseteq W$.
Consider a state $s\in W\cap S_2$, since $U_k\cap W=\emptyset$, 
then $\exists t \in E(s)$ and $t \not\in U_k$ and hence $s \not\in U_{k+1}$.
Consider a state $s \in W \cap S_1$, since $E(s) \subseteq W$ and $W \cap U_k
=\emptyset$, we have $E(s) \cap U_k=\emptyset$. 
Hence $s \not\in U_{k+1}$.
Hence $W\cap U_{k+1}=\emptyset$.

 It follows that $W\cap A_{\wh{s}}=\emptyset$. 
Since $W\subseteq R_{\wh{s}}$ it follows that 
$W \subseteq (R_{\wh{s}}\setminus A_{\wh{s}})$.
The result follows.
\hfill\qed
\end{proof}

\begin{lemma}\label{lemm:improved2}
The total work of step 4.2.2 --- step 4.2.6 of 
Algorithm~\ref{algorithm:improved}  
is $O\big(\frac{n\cdot m}{\log (n)}\big)$ and the total work of step 4.2.7 is 
$O(m)$.
\end{lemma}
\begin{proof}
Consider an iteration of Algorithm~\ref{algorithm:improved}: since 
step~4.2.2 gets executed for $\big(2\cdot\frac{m}{\log (n)}\big)$ 
steps it follows that size of the graph $R_{\wh{s}}$ is 
$O\big(\frac{m}{\log  (n)}\big)$.
It follows that in any iteration the total work of step 4.2.2 --- step 4.2.6 is 
$O\big(\frac{m}{\log (n)}\big)$.
Since there can be at most $O(n)$ iterations of the algorithm the result 
for step 4.2.2 --- step 4.2.6 follows.

 The edges on which step 4.2.7 works are removed for further iteration
when we remove $W_{i+1}$ from the present set of states. 
Hence in step 4.2.7 no edge is worked for more than once. 
Thus we obtain that total work of step 4.2.7 of 
Algorithm~\ref{algorithm:improved} is $O(m)$.
\hfill\qed
\end{proof}

\begin{lemma}\label{lemm:improved3}
The total work in step 4.2.8 of Algorithm~\ref{algorithm:improved} is 
$O\big(n \cdot m \cdot \frac{\log (\delta(G))}{\log(n)}\big)$. 
\end{lemma} 
\begin{proof} In Algorithm~\ref{algorithm:improved} when step 4.2.8 
gets executed let  $\Tr_i$ be the set of vertices identified by the iteration 
of the classical algorithm.
If  $|\Tr_i| \leq \ell-1$, where $\ell$ is the depth of the BFS search of
step 4.2.2,  then it follows from Lemma~\ref{lemm:improved1} that it would have
been identified by step 4.2.7 in of the iteration. Hence every time 4.8 
gets executed at least $\ell$ states are removed from the graph. 
So step 4.2.8 can be executed at most $O\big(\frac{n}{\ell}\big)$ times,
where $\ell \geq O\big(\frac{\log (n)}{\log (\delta(G))}\big)$. 
The work at every iteration is $O(m)$ and hence the total work of step 4.2.8 of
Algorithm~\ref{algorithm:improved} is
$O\big(n \cdot m \cdot \frac{\log (\delta(G))}{\log(n)}\big)$. 
\hfill\qed
\end{proof}

\begin{lemma}\label{lemm:improved4}
The total work in step 4.1 of Algorithm~\ref{algorithm:improved} is 
$O(m \cdot \log (n))$. 
\end{lemma}
\begin{proof}
The condition of step 4.1 ensures that whenever step 4.1 gets executed as 
least $\big(\frac{m}{\log(n)}\big)$ edges are removed from the graph in the
previous iteration.
Hence step 4.1 gets executed at most $\log(n)$ times and each iteration
takes $O(m)$ work. 
The result follows.
\hfill\qed
\end{proof}

Lemma~\ref{lemm:improved2}, Lemma~\ref{lemm:improved3} and 
Lemma~\ref{lemm:improved4} yield the following result.

\begin{theorem}\label{thrm:improved}
The total work of Algorithm~\ref{algorithm:improved} on a game graph $G$ with 
a B\"uchi objective $\Buchi{B}$, where $B \subseteq S$, is 
$O\big(n \cdot m \cdot \frac{\log (\delta(G))}{\log(n)}\big)$. 
\end{theorem}

\begin{remark}
Observe that for a game graph $G$ with $n=|S|$ we have 
$\delta(G) \leq n$, and hence Algorithm~\ref{algorithm:improved} 
is asymptotically no worse than the classical algorithm.
In~\cite{CJH03} an improved algorithm is presented for binary game graphs
(where every state has out-degree at most~2) with a running time of
$O\big(\frac{n^2}{\log(n)}\big)$, for game graphs with $n$-states.
For the special case of binary game graphs the running time of 
Algorithm~\ref{algorithm:improved} matches the bound of the 
algorithm of~\cite{CJH03}.
However, there exists game graphs where Algorithm~\ref{algorithm:improved}
out-performs both the classical algorithm and the algorithm of~\cite{CJH03}.
For example consider the class of game graphs $G$ with $n=|S|$ and 
$|E(s)|=O(\log(n))$ for all states, and hence $m=O(n \cdot \log(n))$.
The classical algorithm and the algorithm of~\cite{CJH03} (after reduction 
to binary game graphs) in the worst case take $O(n^2 \log (n))$ time, 
whereas the worst case running time of Algorithm~\ref{algorithm:improved} is
bounded in $O(n^2 \log (\log (n)))$.
\end{remark}

\end{document}